\documentclass[twocolumns]{aa}
\usepackage{graphicx}
\usepackage{txfonts}
\def\ls{{_<\atop^{\sim}}}
\def\gs{{_>\atop^{\sim}}}

\begin{document}
   \title{\object{Gas reservoir of a hyper-luminous QSO at $\rm z=2.6$}\thanks{
Based on observations carried out with the IRAM Plateau de Bure Interferometer. IRAM is supported by INSU/CNRS (France), MPG (Germany) and IGN (Spain).}
}
%   \subtitle{I. Overviewing the $\kappa$-mechanism}

   \author{C. Feruglio \inst{1,2}
        \and
         	A. Bongiorno \inst{2}
   \and
             F. Fiore\inst{2}
	\and  
        M. Krips \inst{1}	
	\and
        M. Brusa \inst{3,4,5}
        \and
          E. Daddi \inst{6}
         \and 
     	I. Gavignaud \inst{7}
        \and
        R. Maiolino \inst{8}
        \and
	E. Piconcelli \inst{2}
         \and
         M. Sargent \inst{9}
         \and
         C. Vignali \inst{3,5}
         \and 
         L. Zappacosta \inst{2}
 %        R. Maiolino \inst{5}
 %     \and
 %       M. Sargent \inst{4}		
%       \and 
            }
   \institute{IRAM - Institut de RadioAstronomie Millim\'etrique, 
300 rue de la Piscine, 38406 Saint Martin d'H\`eres, France,
              \email{feruglio@iram.fr}
         \and
             INAF - Osservatorio astronomico di Roma, via Frascati 33, 00040 Monteporzio Catone, Italy
	\and 
	University of Bologna, Department of Physics and Astronomy, viale Berti Pichat 6/2, 40127 Bologna, Italy 
	\and 	
	Max Planck Institut f\"ur Extraterrestrische Physik, Giessenbachstrasse 1, 85748 Garching bei M\"unchen, Germany 
	\and
 INAF - Osservatorio Astronomico di Bologna, via Ranzani 1, 40127 Bologna, Italy 
	\and 
	Laboratoire AIM, CEA/DSM-CNRS-Universit\'e Paris Diderot, Irfu/Service d'Astrophysique, CEA Saclay, Orme des Merisiers France, 91191 Gif-sur-Yvette Cedex, France
	\and
	Departamento de Ciencias Fisicas, Universidad Andres Bello, Av. Republica 252, Santiago, Chile 
	\and
	KAVLI
	\and 
	Astronomy Centre, Dept. of Physics and Astronomy, University of Sussex, Falmer, Brighton BN1 9QH, UK
             }

   \date{February 23, 2014}

% \abstract{}{}{}{}{} 
% 5 {} token are mandatory
 
  \abstract 
{Understanding the relationship between the formation and evolution of
  galaxies and their central super massive black holes (SMBH) is one
  of the main topics in extragalactic astrophysics. Links and
  feedback may reciprocally affect both black hole and galaxy growth.}
{Observations of the CO line at redshifts of 2-4 are crucial to
  investigate the gas mass, star formation activity and accretion onto
  SMBHs, as well as the effect of AGN feedback. Potential correlations
  between AGN and host galaxy properties can be highlighted by
  observing extreme objects. Despite their luminosity, hyper-luminous
  QSOs at z$=2-4$ are still little studied at mm wavelengths.}
{We targeted CO(3-2) in ULAS J1539+0557, an hyper-luminos QSO ($\rm
  L_{bol}> 10^{48} erg/s$) at $\rm z=2.658$, selected through its
  unusual red colors in the UKIDSS Large Area Survey (ULAS).}
{We find a molecular gas mass of $\rm 4.1\pm0.8 \times10^{10}~
  M_{\odot}$, and a gas fraction of $\sim$0.4-0.1, depending mostly
  on the assumed source inclination. We also find a robust lower limit to the
  star-formation rate (SFR$=$250-1600 M$_\odot$/yr) and star-formation
  efficiency (SFE$=$25-350 L$_\odot$/(K km s$^{-1}$ pc$^{2}$) by
  comparing the observed optical-near-infrared spectral energy
  distribution with AGN and galaxy templates. The black hole gas
  consumption timescale, $\rm M(H_2)/\dot M_{acc}$, is $\sim 160$ Myr,
  similar or higher than the gas consumption timescale.}
{The gas content and the star formation efficiency are similar to
  those of other high-luminosity, highly obscured QSOs, and at the
  lower end of the star-formation efficiency of unobscured QSOs, in
  line with predictions from AGN-galaxy co-evolutionary scenarios.
  Further measurements of the (sub)-mm continuum in this and similar
  sources are mandatory to obtain a robust observational picture of
  the AGN evolutionary sequence.}


\titlerunning{Gas reservoir of ULAS J1539+0557}
\authorrunning{C. Feruglio}

 \maketitle
 
%
%________________________________________________________________ 

\section{Introduction}

Understanding the relations between the formation and evolution of
galaxies and their central super massive black holes (SMBH) is a major
challenge of present-day astronomy.  Most of galaxy assembly and
accretion activity occur at $\rm z=2-4$, so it is crucial to study
SMBH-galaxy relationships at this epoch.  The two main open questions
are: a) what is the mechanism triggering nuclear accretion and
star-formation? and b) are AGN outflows truly able to regulate
star-formation in their host galaxies?  Observations of molecular gas
are useful to address both questions.  So far, molecular gas has been
detected in a few tens of $\rm z>2$ QSOs, with typical masses
$1-10\times10^{10}$ M$_\odot$, indicating gas-rich hosts. However,
most of these observations have been done either on lensed object
(where the intrinsic luminosity is magnified up to 1000 times), or on
z$=5-6$ QSOs (Riechers 2011 and references therein).

There is a growing evidence for two modes of star-formation which may
also be relevant for triggering nuclear activity: a quiescent one,
taking place in most star-forming galaxies, with gas conversion time
scales of $\sim1$ Gyr, and a less common {\it star-burst} mode, acting
on much shorter time scales ($\sim10^7-10^8$ yr, see e.g. Rodighiero
et al. 2011, Lamastra et al.  2013a and references therein). The
latter mode is likely related to the powering of high luminosity
QSOs. In fact, bolometric luminosities of the order of
$10^{47}-10^{48}$ ergs/s imply mass accretion rates of tens to
hundreds $\rm M_\odot$/yr onto $10^9-10^{10}~\rm M_\odot$ SMBHs
(assuming a radiative efficiency of 0.1).  If the accretion lasts for
a few tens Myr (Salpeter timescale), this in turn implies gas
reservoirs of $10^9-10^{10}$ M$_\odot$ even if the fraction of the gas
that can reach the nucleus is high ($\Delta$M$_{gas}\sim
$M$_{gas}/5$).  At present, galaxy interactions seem to be the best
(if not the only) mechanism capable of destabilizing such huge gas
masses on short time scales. This naturally produces powerful AGN
hosted in star-burst galaxies (Lamastra et al 2013b).  In principle,
from the gas consumption timescale (or its inverse, the so-called star
formation efficiency, SFE, i.e. the ratio between the star-formation
rate and the gas mass), one can directly derive information on the AGN
and star-formation triggering mechanisms.

The SFE depends also on the SMBH gas consumption time scales and on
the energy injected in the ISM by the AGN ({\it feedback}).  High SFE
may be the outcome of a small cold gas reservoir which was reduced by
on-going SMBH accretion and consequent AGN feedback.  For this reason,
to have information on AGN feedback is crucial for addressing both
questions.  There is both theoretical and observational evidence that
high luminosity QSOs drive powerful outflows.
%On the observational ground, significant
%broad components of the [OIII] line were found in about 2000 SDSS QSOs
%(Zhang et al. 2011), confirming that outflows of ionized gas are
%extremely common, in particular at high AGN luminosities. 
On the theoretical ground, physically-motivated models predict strong
winds from AGN with SMBH larger than $10^8$ M$_\odot$(Zubovas \& King
2014 and references therein).  These models in general predict mass
flows proportional to the AGN bolometric luminosity to some power
(e.g. M$_{out}\propto$ L$_{Bol}^{1/2}$ in the Menci et al. 2008
model).  For these reasons, the most luminous QSOs in the Universe are
ideal and unique targets to study AGN/galaxy feedback mechanisms
regulated by powerful outflows.  
Powerful and massive AGN driven outflows of molecular gas were recently discovered in luminous
QSOs (Maiolino et al. 2012, Feruglio et al. 2010, Cicone et al. 2013).
Broad Absorption Lines (BAL) with outflow velocities up to several
thousands km/s are common in high-luminosity QSOs (Borguet et
al. 2013).  BALs are found in 40\% of mid-infrared selected QSOs (Dai et
al. 2008) and in 40\% of the {\it WISE} selected luminous QSOs
(Bongiorno et al. 2014, in prep.).  
Powerful, galaxy wide outflows have also been found in
ionized gas using the broad [OIII] emission line (Cano-Diaz et al. 2012).

%For example outflows of ionized gas, atomic gas and molecular
%gas with kinetic energies $\gs10^{46}$ ergs/s, $\gs10^{45}$ ergs/s,
%and $\gs10^{44}$ ergs/s where discovered in a hyper-luminous
%(L$_{bol}\sim2\times10^{47}$ ergs/s) BALQSO at z=3.04 (Borguet et
%al. 2013), a z$=$6.4 QSO (L$_{bol}\sim10^{47}$ ergs/s, Maiolino et
%al. 2012), and the most luminous QSO in the local Universe (Mark231,
%L$_{bol}\sim 10^{46}$ ergs/s, Feruglio et al. 2010). Compilations of
%recent results, showing that the mass outflow rate seems to increase
%with the luminosity can be found in fCicone et al. (2013) and Fiore et
%al. (2014).  

%%%say some words about unbiased selection
 
In this paper we present results from the first Plateau de Bure
Interferometer (PdBI) observation of an hyper-luminous QSO, selected
from the UKIDSS Large Area Survey (ULAS) and the VISTA (J, K bands)
Hemisphere Survey: ULAS J1539+0557 at $\rm z=2.658$ (ra:15:39:10.2,
dec: 05:57:50.0, Banerji et al. 2012).  ULAS J1539+0557 is heavily
reddened (rest frame $\rm A_V\sim4$) and has a bright 22 $\mu$m flux
of 19 mJy in the{\it WISE} all-sky survey, corresponding to $\lambda
L_{\lambda}(7.8~ \mu$m)$=10^{47}$ ergs/s, and a bolometric luminosity
of $\rm L_{bol}\sim10^{48}$ erg/s.  The black hole mass is as large as
$7.4\times10^{9}~\rm M_{\odot}$.  As detailed in Banerji et
al. (2012a), these hyper-luminous QSOs are unlikely to be lensed,
therefore they truly trace an extremely luminous population.

In the following we present the results of 3 mm observations of ULAS
J1539+0557, targeting the CO(3-2) transition.  We derive the gas mass,
dynamical mass, and star-formation rate (SFR) of the host galaxy
through the comparison of the observed UV-MIR Spectral Energy
Distribution (SED) with galaxy templates.  We compare the observed gas
masses, gas fraction and gas consumption timescale with those of the
other QSOs and galaxies at similar redshift. We finally discuss future
desirable developments in this topic.  A $H_0=70$ km s$^{-1}$
Mpc$^{-1}$, $\Omega_M$=0.3, $\Omega_{\Lambda}=0.7$ cosmology is
adopted throughout.

%__________________________________________________________________

\section{Millimeter observations and results}

ULAS J1539+0557 was observed at a frequency of 94.5 GHz with the PdBI
array in the most compact (D) configuration, in September 2013.  The
system temperature was between 80 and 120 K, and water vapor about 5
mm.  The quasar 1546+027 (1.8 Jy at 94.5 GHz) was used as a phase and
amplitude calibrator.  The quasar 3C454.3 (10.1 Jy) was used for
bandpass and absolute flux calibration.  We estimate a 10\% error on
the absolute flux calibration.  Calibration and mapping were done in
the GILDAS environment.  The flagging of the phase visibilities was
fixed at 35\% rms in order to maximize the signal to noise ratio (S/N) of
the detection.  This flagging yields a 1$\sigma$ sensitivity of 0.3
mJy/beam in 79 MHz channels, and a total 6 antenna equivalent
on-source time of 3.2 hours
%Taking all the data, regardless the phase rms, yields 0.25
%mJy/beam in 79 MHZ channels. 
With natural weighting, the synthesized beam is 8.9 by 5.0 arcsec.  
%\section{Results}

%figura spettro e mappa 
\begin{figure*}
   \centering
    \includegraphics[width=17cm]{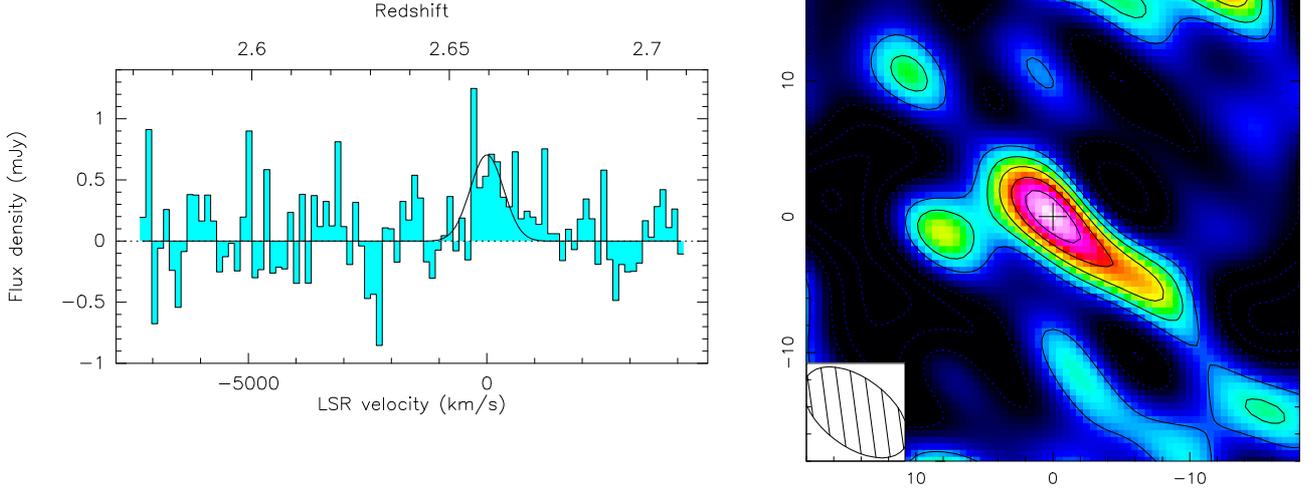}
   \caption{Left panel: spectrum of ULAS J1537+0557 integrated over
     the beam.  The solid line shows the Gaussian fit with FWHM=840$^{+1000}_{-350}$ km/s, and centered
     at the frequency corresponding to the redshift of the source. 
     Right panel: integrated map of CO(3-2). Contour levels are 1$\sigma$ each
     ($\sigma=0.2$ mJy). The synthesized beam is shown in the bottom-left corner.}
\label{co}
\end{figure*}

Fig. \ref{co} shows the spectrum and collapsed map of ULAS J1537+0557.  
The CO(3-2) emission line is detected at a confidence level of 5.4$\sigma$
(0.5,0.3 arcsec offset from the phase center, based on the fit of the
visibilities in the uv plane adopting a unresolved source model) and
lies at the frequency expected based on the optical spectroscopic
redshift. The integrated flux of the line is 1.37 Jy km/s over the
full line width (from fitting of the visibilities by a Gaussian
function).  L$^\prime$CO can be estimated by extrapolating to the
CO(1-0) luminosity from the CO(3-2) flux, using excitation models of
high-z QSOs (Riechers et al. 2006, Riechers 2011).  The observed
scatter (see e.g. Carilli \& Walter 2013) is of the same order of
magnitude as the relative error on the CO flux of ULAS J1539+0557, and
therefore we neglect it in the following estimates. We then find a
line luminosity $\rm L^\prime CO=(5.1\pm1.2)\times 10^{10}$ K km/s
pc$^2$. By using a conversion factor $\rm X_{CO}=0.8$ (e.g. Carilli \&
Walter 2013), we derive a total molecular gas mass of $\rm M(H_2)=
(4.1\pm0.8)\times 10^{10}~ M_{\odot}$.  The 3 mm continuum is not
detected (with a 3$\sigma$ upper limit of 0.1 mJy).

The CO(3-2) line profile in Fig. \ref{co} is broad. Fitting 
the profile over the full velocity range with a single Gaussian
component we obtain a FWHM is 1600$\pm700$ km/s. 
This would be the highest FWHM among high-z QSOs (average is $\sim300$ km/s, Riechers 2011).  
Although noisy, the line profile in Fig. \ref{co} appears asymmetric, skewed toward positive velocities. 
The feature at positive velocities might either be a different component of the emission or noise.
If we exclude from the fit  velocities $>500$ km/s the best fit
line width is narrower, FWHM=840$^{+1000}_{-350}$ km/s.
Fitting the profile with two Gaussian functions we obtain a best fit
solution with a main component centered at zero velocity with
FWHM=1100$\pm$450 km/s and a second, fainter component centered at
about 1200 km/s. 
This would suggest that the complex line profile is the result of a merging system (see e.g. Fu et al. 2013). 
The S/N of the data is not good enough to disentangle between the different possibilities. 
In the following we conservatively assume that the FWHM of the line is 840$^{+1000}_{-350}$ km/s.

The source appears unresolved, but the limits on its extension are
quite loose (physical size $<40$ kpc) because of the large synthesized
beam.

The CO line width can be converted into a dynamical mass assuming a
size $\rm R$ and an inclination $i$ of a rotating molecular gas disc.
We derive the product of the circular velocity at the outer CO radius
$v_c$ times the sinus of inclination, $v\times sin(i)$, by dividing
the FWHM of the CO line by 2.4 (Tacconi et al. 2006). The dynamical
mass can then be estimated as $\rm M_{dyn}~sin^2(i)=R~\Delta
v^2_{c}/G=5.7\times10^{10}$M$_\odot$, assuming $\rm R=2~ kpc$, a value
commonly used for QSOs (e.g. Coppin et al. 2008), similar to that
measured for the molecular disc of the z=4.694 southern AGN of
BR1202-0725 (Carniani et al. 2013), or the lower luminosity QSOs
(Krips et al. 2007), or local QSOs (e.g. Mark231, Downes \& Solomon
1998), and similar to that found in SMGs (Tacconi et al. 2006, 2008).

Table 1 compares CO line FWHM, gas and dynamical mass of ULAS
J1539+0557 to those of the only other three available $\rm z>2$,
hyper-luminous ($\lambda L_{\lambda} (7.8~\mu$m)$\gs10^{47}$ ergs/s)
QSOs.

The large FWHM of CO(3-2) of ULAS J1539+0557 suggests a non negligible
inclination. If $i=20$ deg, as in SBSJ1408+567 (Coppin et al. 2008),
M$_{dyn}~\sim 5\times10^{11}~ \rm M_\odot$, if i=45 deg, M$_{dyn}~\sim
10^{11}~ \rm M_\odot$.  The ratio between the gas mass and the
dynamical mass $M_{H2}/M_{dyn}$ is therefore likely in the range
0.4-0.1.

\section{Spectral Energy Distribution}

\begin{figure}
   \centering
   \includegraphics[width=\columnwidth]{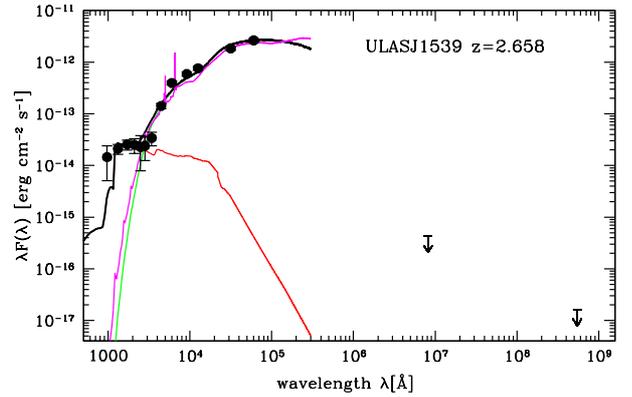}
   \caption{The optical to radio wavelength rest-frame SED of ULAS
     J1539+0557 fitted with 2 components: a highly extincted AGN model
     (E(B-V)=1.1, magenta line), and a galaxy component (with
     different extinction, E(B-V)=0.2, red line). The black solid line shows the 
     sum of the two components. 
     The arrows indicate the $3\sigma$ upper limits in the continuum at 3 mm (these
     observations) and at 20 cm from the FIRST survey.}
\label{sed}
\end{figure}

ULAS J1539+0557 has never been observed at far infrared or sub-mm
wavelengths. It is possible, however, to compile a robust broad-band
spectral energy distribution (SED) from the UV to the 22 $\mu$m band
(photometry from Banerji et al. 2012).  The source is undetected in
the FIRST VLA survey (Becker et al. 2012), the 3$\sigma$ upper limit
is 0.375 mJy at 20 cm.  The SED in the QSO rest frame is shown in
Fig. \ref{sed}.  The SED at rest-frame wavelengths above $\sim4000$
$\AA$ is clearly dominated by the AGN.
The emission of the AGN rapidly drops below 3000 $\AA$, indicating
that the nuclear emission is substantially obscured, consistent with
the color selection of this source in the VISTA survey (Banerji et
al. 2012).  The SED below 3000 $\AA$ is likely dominated by starlight
from the host galaxy.  Translating the observed 1500 \AA\ luminosity
into a SFR, assuming the Madau (1998) conversion and zero galaxy
extinction, provides a firm lower limit to the host galaxy SFR of 70
M$_\odot$/yr.  The real SFR is probably much higher than this value,
if the galaxy is substantially obscured by dust, but difficult to
constrain due to the lack of far-IR data.

To obtain a better SFR estimate, we modeled the broad band, 0.1-6
$\mu$m SED by using a library of AGN and galaxy templates (Bongiorno
et al. 2012).  For the AGN component, we used the mean QSO SED from
Richards et al.  (2006), while for the galaxy component, a library of
synthetic spectra generated using the stellar population synthesis
models of Bruzual \& Charlot (2003) has been adopted. Both the AGN
and the galaxy templates can be affected by dust extinction.  For a
given galaxy template, the free parameters in the fit are thus
normalization and extinction of both AGN and galaxy templates.  We
accepted solutions with galaxy stellar masses smaller than $10^{12}$
M$_\odot$ and SFR $\times $ duration of the starburst $\ls$ total
stellar mass. 
%bearing in mind that the data do not allow a robust estimate of the stellar mass.  
The best fit values are 400 M$_\odot$/yr of SFR 
and $3\times10^{10}~\rm M_{\odot}$ for the stellar mass.
The 1$\sigma$ confidence intervals for SFR and stellar mass are
250-1600 M$_\odot$/yr and $3\times10^{10}-3\times10^{11}$ M$_\odot$
respectively.  The best fit extinction for the AGN and the galaxy is
$\rm E(B-V)=1.1$ and 0.2, respectively.

%figura SED

\begin{table*}
\caption{\bf Hyper-luminous QSOs at $\rm z>2$ }
\centering
\begin{tabular}{lcccccccc}
\hline
QSO              & z      &  $\lambda L_{\lambda} (7.8~\mu$m) & M(H$_{2})$ & FWHM  & M$_{dyn}sin^2(i)$  &  M$_{H2}$/M$_{dyn}$ & M$_{BH}$ & Ref \\
                 &        &  [$10^{47}$ ergs/s]              & [$10^{10}$M$_\odot$] &  [km/s] & [$10^{10}$ M$_\odot$]  &                   & [$10^{9}$ M$_\odot$]  \\
\hline
ULAS J1539+0557   & 2.658  & 1.0 & 4.1$\pm0.8$ & 840$^{+1000}_{-350}$ (3-2)  & 5.7$^a$ & 0.4-0.1$^d$  & 7.4 & (1) \\ 
\hline
RXJ1249-0559     & 2.247  & 1.7 & 2.9$\pm0.8$ & 1090$\pm340$ (3-2)         & 9.7$^a$ & 0.30$^b$  & 5.7 & (2) \\
SBSJ1408+567     & 2.583  & 1.8 & 6.0$\pm0.5$ & 311$\pm28$  (3-2)          & 0.78$^a$& 0.90$^c$  & 1.1 & (3) \\
BR1202-0725      & 4.694  & 1.6 & 3.2$\pm1.2$ & 363$\pm37$  (5-4)          & 0.83    & 0.25$^e$ & 1.5 & (4,5) \\
\hline
\end{tabular}

$^a$assuming a disk radius of 2 kpc;
(1) Banerji et al. 2012;
(2) Coppin et al. 2008, 
(3) Beelen et al. 2004; 
(4) Salome et al. 2012;
(5) Carniani et al. 2013, 
$^b$ likely high inclination;
$^c$ assuming $i=20$ deg;
$^d$ assuming $i=45$ deg and $i=20$deg;
$^e$ assuming $i=15$ deg.
\end{table*}

\section{Discussion}

ULAS J1539+0557 is an extreme object, with a bolometric luminosity of
$\rm L_{bol}=1.6\times 10^{48}$ erg/s (derived from the 5100 \AA~
continuum), and a SMBH mass, estimated from H$\alpha$ emission line,
as large as $7.4\times10^{9}~\rm M_{\odot}$ (Banerji et al. 2012).
This implies that the SMBH is accreting at the Eddington limit, at a
rate of $dm/dt = 200-300 \rm ~M_{\odot}/yr$.  At this rate, the black
hole gas consumption timescale, $\rm M(H_2)/\dot M_{acc} $, would be
only $\sim 160 ~\rm Myr$.

In the following we discuss the source properties in comparison with other
known luminous systems.  A clean estimator for the intrinsic AGN power
is its mid-infrared luminosity, since it guarantees little bias
against dusty objects and little extinction (or absorption)
corrections.  Recently, the {\it WISE} satellite has performed the
deepest all-sky survey at 22 $\mu$m, which samples the rest frame 4-8
$\mu$m band at $\rm z=2-4$ (where AGN heated warm dust is the main
component to the total flux), finding the most luminous AGN in the sky at z$<4-5$.  
It is therefore convenient
to use the 7.8 $\mu$m luminosity ($\lambda L_{\lambda}(7.8~ \mu$m)) as
a reliable proxy for the AGN power.  By measuring $\lambda
L_{\lambda}(7.8~\mu$m) of all QSOs with CO detections, we find that
only three of these sources at $\rm z<5$ have extreme intrinsic
luminosities ($\lambda L_{\lambda}(7.8~ \mu$m)$\gs10^{47}$ ergs/s).
Other six QSOs with CO detection have $10^{46}<\lambda
L_{\lambda}(7.8~\mu$m)$<10^{46.5}$ ergs/s, including the two highly
obscured SWIRE QSOs of Polletta et al. (2011).  To characterize
quantitatively the unique population of hyper-luminous QSOs we need to
substantially increase the sample with gas mass, dynamical mass and
SFR detections.

\begin{figure}
   \centering
   \includegraphics[width=8.5cm]{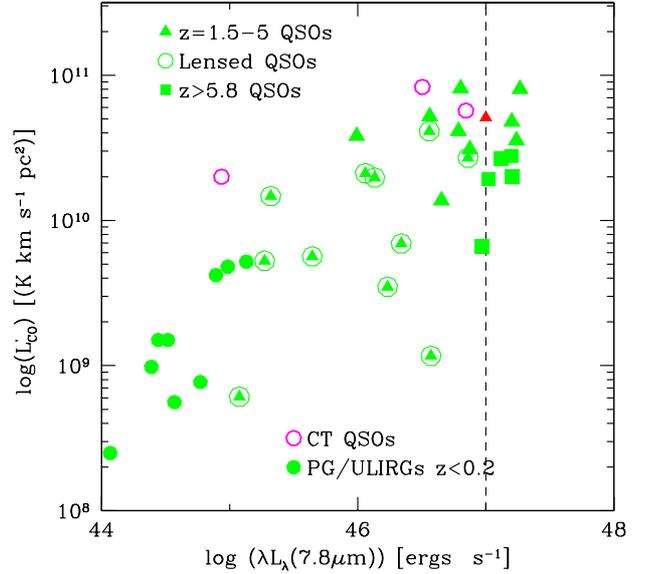}
   \caption{$\rm L^\prime CO$ versus $\lambda L_{\lambda} (7.8\mu$m) for a 
compilation of AGN. The red triangle corresponds to ULAS J1539+0557.}
\label{lcol78}
\end{figure}

Fig. \ref{lcol78} plots $\rm L^\prime CO$ against $\lambda L_{\lambda}
(7.8~\mu$m) for a compilation of both unobscured and highly obscured AGN.  
The luminosities of lensed object have been corrected for the published
lens amplification factor.  A rough correlation between the AGN
luminosity and the CO luminosity is present.  
Note that the scatter in the correlation is higher for the lensed objects, likely due to the
fact that both luminosities have been corrected for the same
amplification factor, but the source of the 7.8 $\mu$m luminosity is
most likely very compact (pc scale), while the source of the CO
luminosity could be more extended (a few kpc), implying an
amplification pattern more complex than the assumed one.  
The 7.8 $\mu$m luminosities of highly obscured (Compton Thick, $\rm N_H>10^{24}~cm^{-2}$, 
two are from Polletta et al. 2011, and Mrk231) have been computed 
using the 6.5-22 $\mu$m fluxes, not corrected for extinction,
which however may be relevant in highly obscured objects even at such
long wavelengths.  
AGN at z$>5.8$ are found at the lower part of the correlation.  
Non-lensed QSOs at z$=1.5-5$ with $\rm \lambda L_{\lambda}
(7.8~\mu$m)$>$a few$10^{46}$ ergs/s have all high $\rm L^\prime CO$,
in the range $3-8\times10^{10}$ K km/s pc$^2$.

%figura 

To convert $\rm L^\prime CO$ into gas mass we again adopt a conversion
factor $\rm X_{CO}=0.8$, as in most studies of AGN, ULIRGs and SMGs
(see Carilli \& Walter 2013).  The molecular gas mass of ULAS
J1539+0557 and three other hyper-luminous QSOs (Table 1) are in the
range 3-6$\times10^{10}$ M$_\odot$.  We evaluate the dynamical mass of
ULAS J1539+0557 under admittedly major assumptions (the FWHM estimated
from the core of the CO(3-2) line, the size of the molecular gas disc
of 2 kpc and its inclination angle).  We find that the gas mass is 0.1-0.4 of the dynamical mass.  
This suggests that the host galaxy of this hyper-luminous
QSO is not deprived from gas, and it is likely forming
stars actively, in agreement with the prediction of semi-analytic
models for AGN/galaxy co-evolution (Lamastra et al. 2013b).

%{\bf The small M$_{H2}$/M$_{dyn}$ ratio might also indicate that AGN
%  feedback has already been able to significantly modify the
%  environment.  In that case the SFE may not be a reliable indicator
%  of the gas consumption timescale.  } 

SFE is usually computed as $\rm SFE=L_{FIR}$/L$^\prime$CO.  
Since ULAS J1539+0557 has never been observed at far infrared (FIR) or sub-mm
wavelengths, we do not have a direct measure of L$_{FIR}$. We can,
however, estimate L$_{FIR}$ from the SFR obtained from the SED
fitting, using the conversion SFR=$2-1.2\times
10^{-10}$(L$_{FIR}$/L$_\odot$) M$_\odot$/yr (Scoville 2012).  We find
SFE between 25 and 350 L$_\odot$/(K km s$^{-1}$ pc$^{2}$), which
correspond to a gas consumption timescale of 25-160 Myr, smaller or
equal to the black hole gas consumption timescale computed above
(assuming that AGN feedback has not yet been able to significantly
modify its environment and reduce the total host galaxy cold gas
mass).

For a SFR in the range 250-1600, the ratio $\rm \dot M_{acc}/SFR$ is
in the range $0.1-1$, which is $\sim2-3$ orders of magnitude above the
mean value $10^{-3}$ found by Mullaney et al. (2012) for AGN with
star-forming hosts in GOODS-S, thus suggesting that ULAS J1539+0557 is
an outlier to the "AGN main sequence".  The SMBH mass is $\gs2\%$ of
the dynamical mass, so $M_{BH}/M_*$ is of the order $0.1$ or higher,
i.e. 2 orders of magnitude larger than the canonical value.  This
means that, although ULAS J1539+0557 has a extremely massive SMBH,
this seems to live in an ordinary galaxy host with $\rm
M_{dyn}\sim10^{11}~M_{\odot}$, which is uncommon at this redshift, but not as such at higher redshift.  
$\prime$Stochastic fluctuations$\prime$ between SMBH and SF activity, as advocated by
Mullaney et al. (2012), are unlikely to give rise to such extreme
outliers.

\begin{figure*}
   \centering
\begin{tabular}cc
   \includegraphics[width=8.5cm]{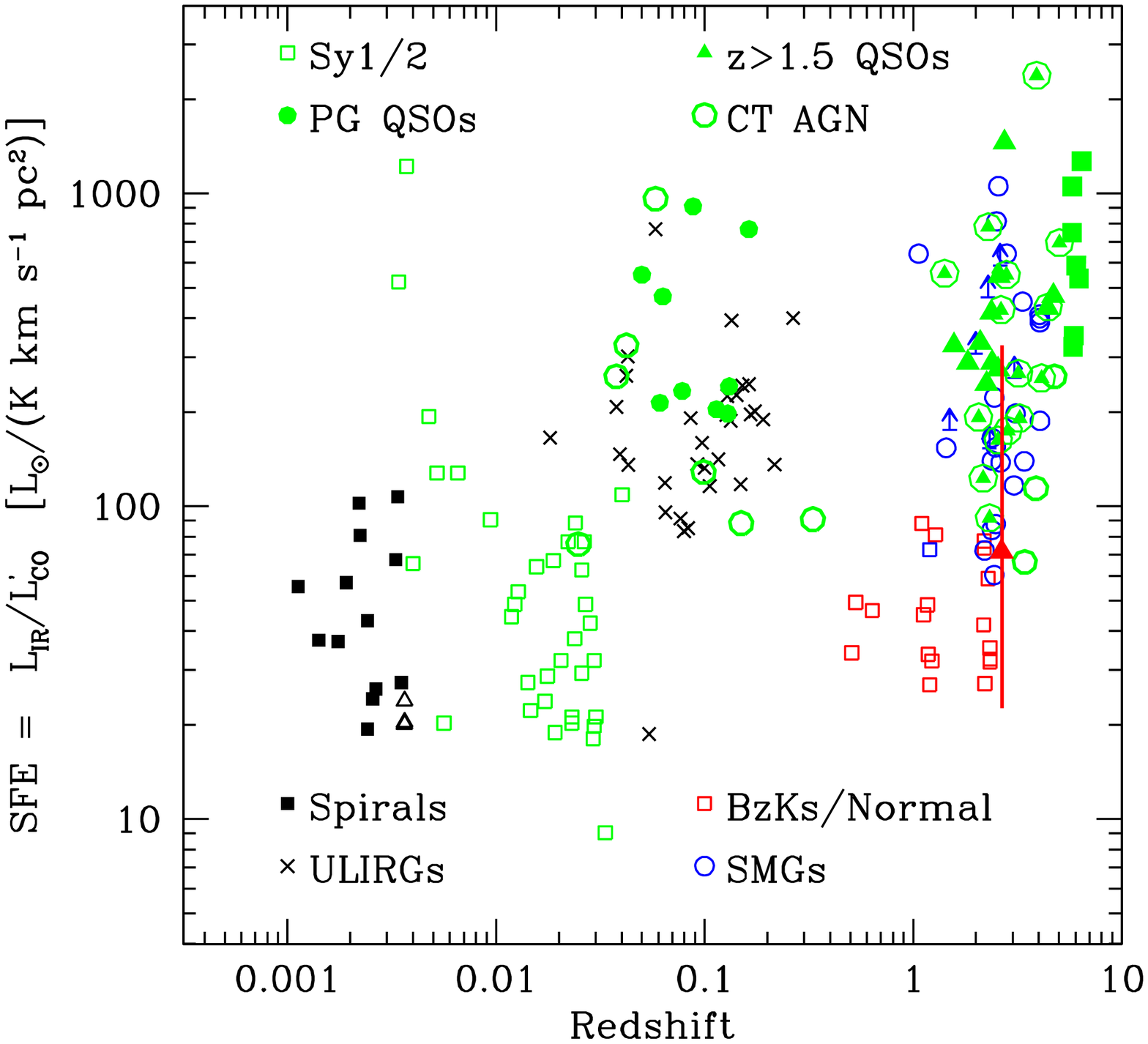}
   \includegraphics[width=8.5cm]{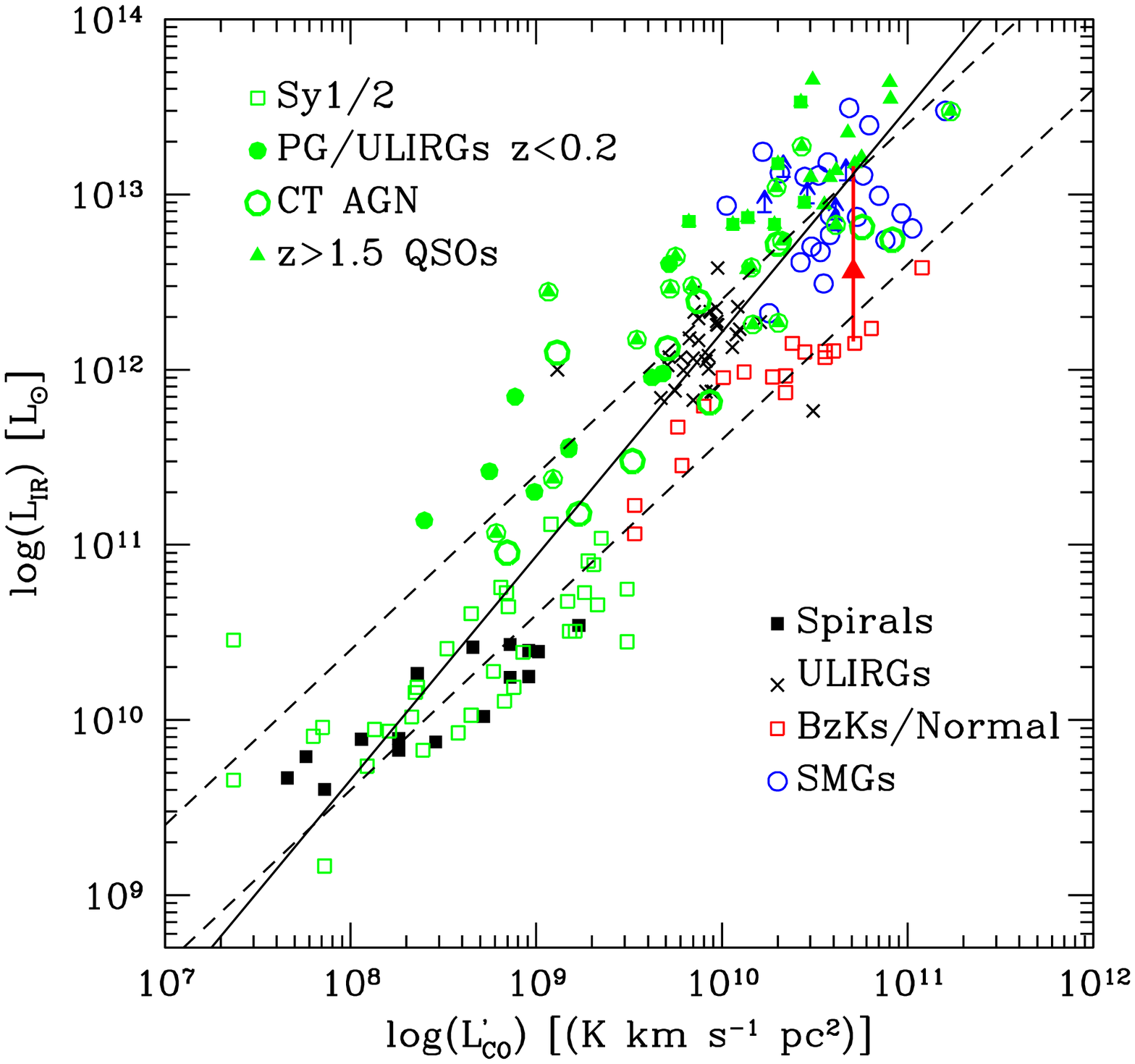}
\end{tabular}
   \caption{[Left panel]: SFE versus redshift for a compilation of AGN
     and star-forming galaxies (from Daddi et al. 2007, Genzel et
     al. 2010, Riechers et al. 2011 and references therein, Krips et
     al. 2012).  The red triangle corresponds to ULAS
     J1539+0557. [Right panel]: L$_{FIR}$ versus $\rm L^\prime CO$ for
     a compilation of AGN and star-forming galaxies at all
     redshifts. The red triangle corresponds to ULAS J1539+0557. The
     solid line is a fit to all data, the dashed lines are the best
     fit for main sequence galaxies and star-burst galaxies (Daddi et
     al. 2010, Genzel et al. 2010).}
\label{lirlco}
\end{figure*}

In Fig. \ref{lirlco} we plot the SFE against redshift (left panel) 
and L$_{FIR}$ against $\rm L^\prime CO$ (right panel), for a compilation of AGN and
galaxies (samples of Daddi et al. 2007, Genzel et al. 2000, Riechers
et al. 2011 and references therein, Krips et al. 2012). 
The uncertainty on L$_{FIR}$ and SFE, due to both the uncertainty on the
evaluation of the SFR from SED fitting, and that associated to the
conversion from SFR to L$_{FIR}$, prevents us from precisely locating
ULAS J1539+0557 in the L$_{FIR}$--$\rm L^\prime CO$ plane.  Its FIR
luminosity is consistent with both the locus of normal, star-forming
galaxies and that of SMGs, ULIRGs and luminous QSOs.  At face value,
the SFE of ULAS J1539+0557 is lower than (although still statistically
consistent) that of other luminous QSOs at similar or lower redshift
(Riechers et al. 2011).  It is similar to that of other highly
obscured, but less extreme AGN.  For example, Krips et al. (2012)
found SFE$\sim100-200$ in nearby type 2 QSOs, and Polletta et
al. (2011) found similar SFE in two galaxies hosting high-luminosity,
obscured QSOs at $\rm z\sim3$.  Therefore ULAS J1539+0557 may be an
extreme case of a dust enshrouded, hyper-luminous QSO.

In the AGN/galaxy coevolution scenario obscured AGN may be an early
phase of the life cycle of luminous quasars (Sanders et al. 1988,
Menci et al. 2008 and references therein), in which the strong AGN
outflows have not yet cleared the host galaxy from most of its gas and
dust, thus resulting in a lower SFE than that of unobscured quasars.
Conversely, luminous unobscured QSOs may be caught at the end of this
process, and have high SFE because the AGN feedback has been already
efficient in expelling a large fraction of the gas.  The uncertainties
on the SFR and SFE of ULAS J1539+0557 are, however, too large to allow
a firm conclusion.

%figura 

\section{Conclusions}

We detected CO(3-2) in the hyper-luminous QSO ULAS J1539+0557 at $\rm
z=2.658$, a source selected by extremely red colors in the UKIDSS
Large Area Survey (ULAS) and bright at 22 micron in the {\it WISE}
survey.  We find a molecular gas reservoir of $\rm 4.1\pm0.8 ~10^{10}~
M_{\odot}$. The dynamical mass is not well constrained but it could be
2-10 times higher, depending mostly on the gas disk size and
inclination. In any case, the host galaxy does not appear deprived
from cold gas, suggesting that it is still forming stars actively.
From fitting of the UV to mid-IR SED, we derived a robust lower limit
to the SFR and gas consumption time scale.  
The ratio of SMBH accretion to star-formation rate, $\rm \dot M_{acc}/SFR$, is 
significantly higher than that found by Mullaney et al. (2012) for AGN with star-forming
hosts.  The SFE is similar to that of highly obscured, high luminosity
QSOs.  This class of QSOs is believed to witness the brief
evolutionary phase that traces the transition from a heavily
enshrouded ULIRG-like phase of black-hole growth to the blue
unobscured quasars.  Due to their high luminosity these exceptional
objects are ideal laboratories to investigate the physics of the
feedback phenomenon with in situ observations at the peak of galaxy
and SMBH assembly.  The high luminosity allows highlighting the
correlation between SFE and other parameters (e.g. obscuration, AGN
luminosity).
 
Both the {\it WISE} all sky survey and the UKIDSS and VISTA hemisphere
Surveys have revealed a population of hyper-luminous QSOs in near and
mid infrared.  These selections have already produced reasonably large
hyper-luminous QSO samples at z$\sim1.5-4$, the main epoch of galaxy
formation and accretion activity in the Universe. 
Recently, Banerji et al. (2014) presented IR and X-ray observations of
a similar source from the same parent sample, ULAS J1234+0907 at
redshift 2.5, finding high SFR ($\gs 2000\rm M_\odot/yr$) and high
X-ray luminosity ($\gs 10^{45}$ergs/s), thus confirming that infrared
selection is efficient in discovering a population of hyper-luminous
QSOs in the {\it blowout} phase.

It is urgent to increase the sample of hyper-luminous QSOs with good estimates of both
gas mass, dynamical mass, L$_{FIR}$, and X-ray luminosity, including both unobscured and
highly obscured QSOs.  These goals can be achieved with the present
and future generation of millimeter interferometers, such as the PdBI,
NOEMA and ALMA.

\begin{acknowledgements}
FF, AB and CF acknowledge support from PRIN-INAF 2011. MB acknowledges
support from the FP7 Career Integration Grant "eEASy" (CIG 321913).
IG acknowledges support from FONDECYT through grant 11110501.
\end{acknowledgements}

%%%%NOTE CO(3-2) = 9 CO(2-1) in thermal equilibrium (opt thin). This is the maximum value. 

\end{document}